\documentclass[preprint,jgrga]{agutex}
  \usepackage[dvips]{graphicx}

\authorrunninghead{HATHAWAY \& UPTON}

\titlerunninghead{MERIDIONAL FLOW AND THE SUNSPOT CYCLE}

\authoraddr{Corresponding author: D. H. Hathaway,
Mail Code ZP13, NASA/Marshall Space Flight Center, Huntsville, AL 35812 USA.
(david.hathaway@nasa.gov)}

\begin{document}

%% ------------------------------------------------------------------------ %%
%
%  TITLE
%
%% ------------------------------------------------------------------------ %%

\title{The Solar Meridional Circulation and Sunspot Cycle Variability}

%% ------------------------------------------------------------------------ %%
%
%  AUTHORS AND AFFILIATIONS
%
%% ------------------------------------------------------------------------ %%

\authors{D. H. Hathaway\altaffilmark{1}
  and L. Upton\altaffilmark{2,3}}

\altaffiltext{1}{Space Science Office, NASA Marshall Space Flight Center,
Huntsville, Alabama, USA.}

\altaffiltext{2}{Department of Physics and Astronomy,
Vanderbilt University, Nashville, Tennessee, USA.}

\altaffiltext{3}{Center for Space Physics and Aeronomy Research,
University of Alabama in Huntsville, Huntsville, Alabama, USA.}

%% ------------------------------------------------------------------------ %%
%
%  ABSTRACT
%
%% ------------------------------------------------------------------------ %%

\begin{abstract}
We have measured the meridional motions of the magnetic elements in the Sun's surface layers since 1996 and find systematic and substantial variations.
In general the meridional flow speed is fast at cycle minima and slow at cycle maxima.
We find that these systematic variations are characterized by a weakening of the meridional flow on the poleward sides of the active (sunspot) latitudes.
This can be interpreted as a inflow toward the sunspot zones superimposed on a more general poleward meridional flow profile.
We also find variations in the meridional flow which vary from cycle-to-cycle.
The meridional flow was slower at both the minimum and maximum of cycle 23 compared to similar phases of cycles 21, 22, and 24.
Models of the magnetic flux transport by a variable meridional flow suggest that it can significantly modulate the size and timing of the following sunspot cycle through its impact on the Sun's polar magnetic fields.
We suggest that the meridional flow variations observed in cycle 23 contributed to the weak polar fields at the end of the cycle which then produced a weak cycle 24 and the extraordinary cycle 23/24 minimum.
\end{abstract}

%% ------------------------------------------------------------------------ %%
%
%  BEGIN ARTICLE
%
%% ------------------------------------------------------------------------ %%

% The body of the article must start with a \begin{article} command
%
% \end{article} must follow the references section, before the figures
%  and tables.

\begin{article}

%% ------------------------------------------------------------------------ %%
%
%  TEXT
%
%% ------------------------------------------------------------------------ %%

\section{Introduction}

The sunspot cycle minimum between cycles 23 and 24 (cycle 23/24 minimum) was exceptional compared to others in modern times.
In December of 2008 the 13-month smoothed sunspot number reached its lowest level since July of 1913 and the smoothed number of spotless days in a month reached its highest level since August of 1913.
In September of 2009 geomagnetic activity, as measured by the \textit{aa} index, reached record lows (since measurements began in 1868) while galactic cosmic rays, as measured by neutron monitors, reached record highs (since measurements began in 1953).

Since cycle 23/24 minimum in late-2008 we have seen the rise of cycle 24 as the smallest sunspot cycle in 100 years.
This provides the simple answer to the question: What caused this extraordinary sunspot cycle minimum?
This deep and extended minimum was caused by the typically delayed start of a small cycle.
Statistically, small cycles start late and leave behind a long cycle and a deep minimum \citep{Hathaway_etal99} (c.f. minima preceding cycles 12-15).
Two effects contribute to this: the actual start times for small cycles are delayed \citep{Hathaway_etal94} and the Waldmeier Effect \citep{Waldmeier35} (in which small cycles rise more slowly) moves the date of cycle minimum between overlapping cycles.

This explanation raises an obvious follow-on question: What caused cycle 24 to be so small?
This can be attributed to the weak polar fields built up during cycle 23.
Models of the Sun's magnetic dynamo suggest that the Sun's largely dipole magnetic field at cycle minimum is the seed for the magnetic field that erupts in the form of sunspots after amplification by the Sun's differential rotation \citep{Babcock61, Leighton69}.

Based on these models, \citet{Schatten_etal78} proposed that the strength of the polar fields at sunspot cycle minimum should be a good predictor of the amplitude of the following sunspot cycle maximum.
Recently, \citet{MunozJaramillo_etal13A} confirmed (by using counts of polar faculae as a proxy for the polar fields prior to 1976) that the polar fields themselves are indeed well correlated with the amplitude of the following sunspot cycle maximum.
Sunspot cycle predictions based on the polar fields at minimum have successfully predicted the last three cycles \citep{Schatten_etal78, SchattenSofia87, Schatten_etal96} as well as cycle 24 \citep{Svalgaard_etal05}.

\citet{WangSheeley09} have noted that the strength of Sun's axial dipole is more closely attuned to dynamo theory and may be measured more accurately than the polar fields for previous cycles.
The axial dipole largely determines the interplanetary magnetic field near cycle minima and 
this field can be derived from historical geomagnetic measurements \citep{SvalgaardCliver05, Rouillard_etal07}.
In fact, \citet{WangSheeley09} suggest that it is this connection that makes the geomagnetic \textit{aa} index at its minimum such a good predictor for the amplitude of following cycle \citep{Ohl66, Hathaway_etal99}.

This explanation for the cause of a weak cycle 24 minimum raises yet another follow-on question: Why did cycle 23 produce such weak polar fields and axial dipole?
Dynamo models \citep{Babcock61, Leighton69} and models for the transport of magnetic flux in the solar photosphere \citep{DeVore_etal84, Wang_etal89, vanBallegooijen_etal98, SchrijverTitle01, Baumann_etal04} produce the polar fields through the emergence of tilted bipolar active regions followed by the poleward transport of the magnetic flux.
The strength of the polar fields depends on the details of both the sources (the total magnetic flux and tilt of the active regions) and the transport processes (the meridional flow and the non-axisymmetric cellular convective flows).
Either the sources or the transport processes, or both, must vary to give variable polar fields and the consequent solar cycle variability.

\section{Producing the Sun's Axial Dipole}

\begin{figure}[ht!]
\centerline{\includegraphics[width=\columnwidth]{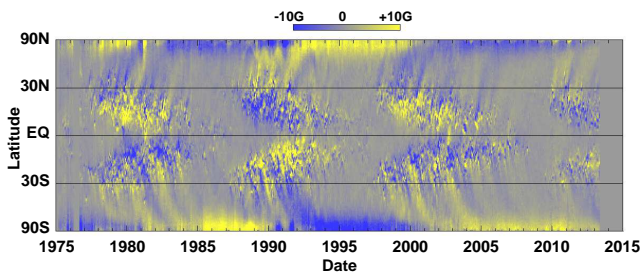}}
\caption{A magnetic butterfly diagram produced from data acquired by the National Solar Observatory. This shows the emergence of magnetic flux with the systematic separation of polarities in the active latitude bands as well as the poleward transport of this flux.}
\label{fig:magbfly}
\end{figure}

Both the active region sources and the transport processes that produce the axial dipole moment are evident in magnetic butterfly diagrams (Fig.~\ref{fig:magbfly}).
These diagrams are produced by averaging the Sun's photospheric magnetic field over longitude during each solar rotation to show the magnetic field as a function of latitude and time.
The butterfly wings in these diagrams show the emergence of magnetic flux in the active latitudes with a predominance of one polarity at higher latitudes and the opposite polarity at lower latitudes due to the Joy's Law tilt of active regions \citep{Hale_etal19}.
This polarity pattern alternates from hemisphere-to-hemisphere and cycle-to-cycle reflecting Hale's Law \citep{Hale_etal19}.

As each cycle progresses, new tilted dipoles emerge at progressively lower latitudes so that the higher latitude following polarity flux progressively cancels the preceding polarity flux that previously occupied those latitudes.
The poleward transport of the high latitude polarity is evident in Fig.~\ref{fig:magbfly} in the form of unipolar streams that move poleward with time.
As each cycle reaches its maximum phase the sunspot zones extend to their lowest latitudes with leading polarity in close proximity to the equator.
This allows for the diffusive transport of leading polarity flux across the equator and cancellation with the (opposite polarity) leading polarity flux in the other hemisphere.

\begin{figure}[ht!]
\centerline{\includegraphics[width=\columnwidth]{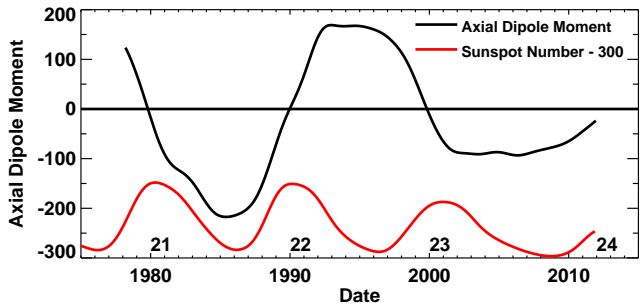}}
\caption{The Sun's axial dipole moment (black line) as measured at the Wilcox Solar Observatory for the last 40 years.
The smoothed sunspot number is shown in red.
This shows that the dipole reverses at about the time of cycle maximum and reaches its maximum at about the time of cycle minimum.
The amplitudes of the following cycles decrease as the amplitudes of the dipole moment decrease.}
\label{fig:AxialDipole}
\end{figure}

The polarity of the high latitude flux is opposite to that of the polar fields at the start of the cycle so the emergence and poleward transport cancels the polar fields from the previous cycle and then builds up opposite polarity polar fields over the remainder of the sunspot cycle.
This is seen in the axial dipole moment (Fig.~\ref{fig:AxialDipole}) calculated from the latitudinal distribution of the surface magnetic field.
The dipole reversals occur at about the time of cycle maximum and the axial dipole moment that is built up at the end of the cycle is well correlated with the strength of the following cycle
\citep{WangSheeley09, MunozJaramillo_etal13A}.

This process could potentially lead to catastrophic behavior for solar cycle amplitudes.
Big sunspot cycles have more magnetic flux emerging in active regions which would produce stronger axial dipoles along with increasing cycle amplitudes.
Likewise, smaller sunspot cycles should produce a string of cycles with diminishing amplitudes.
While the sunspot record does show periods of increasing and decreasing cycle amplitudes, these trends are always limited to 4-5 cycles as part of the 100-year Gleissberg Cycle \citep{Gleissberg39}.
This suggests that either the transport processes or the Joy's Law tilt of active regions (or both) must somehow change systematically with sunspot cycle amplitude to stem the tide of increasing or decreasing cycle amplitudes.

A study of active region tilt by \citet{DasiEspuig_etal10} indicates that the tilt may indeed vary with sunspot cycle amplitude.
They found a tendency for a smaller proportionality between active region tilt and latitude in bigger cycles.
This effect would help to modulate the polar field production by leaving less of an excess of the higher latitude polarity during big cycles as shown by \citet{Cameron_etal10} and \citet{Jiang_etal11A}.

Another mechanism for modulating the polar fields is to modulate the latitudinal transport.
This transport is facilitated by two processes -- a random walk by the rapidly evolving convection pattern and the direct transport by the poleward meridional flow.
The convection pattern includes granules with velocities of $\sim3000$ m s$^{-1}$, lifetimes of $\sim10$ minutes, and diameters of $\sim1$ Mm and supergranules with velocities of $\sim300$ m s$^{-1}$, lifetimes of a day, and diameters of $\sim30$ Mm.
In one year, a random walk by supergranules should have $\sim400$ ``steps'' of ~15 Mm giving a displacement of $\sim300$ Mm
The meridional flow velocity is only $\sim10$ m s$^{-1}$ but this direct flow also gives a displacement of $\sim300$ Mm over the course of a year.
While these simple calculations suggest that these two processes have similar impact, it should be noted that the diffusive effect of the supergranules gives both poleward and equatorward motions and the magnitude of the diffusivity still remains uncertain \citep{Hagenaar_etal99}.

In this paper we report on our measurements of the meridional motions of the magnetic elements.
We find variations in the strength and structure of the meridional flow that may help to explain in general how the polar fields are modulated and in particular how they were modulated during cycle 23.
These variations may have contributed to the production of the weak axial dipole moment during cycle 23 which, in turn, caused the small amplitude for cycle 24.

\section{Measuring the Meridional Flow}

Measurements of the meridional flow can be made by a variety of techniques including direct Doppler, local helioseismology, and feature tracking.
Advantages and disadvantages of the different techniques are discussed in the Appendix.
We choose to use magnetic feature tracking because is not masked by large systematic signals and it measures the motions of the features of interest for magnetic flux transport -- magnetic field elements.

Our measurement technique is described in more detail in \citet{HathawayRightmire10}, \citet{HathawayRightmire11}, and \citet{RightmireUpton_etal12}.
We acquired full-disk magnetograms from the \textit{ESA/NASA Solar and Heliospheric Observatory} Michelson Doppler Investigation (SOHO/MDI) from May 1996 through March 2011 and from the \textit{NASA Solar Dynamics Observatory} Helioseismic and Magnetic Imager (SDO/HMI) from April 2010 through July 2013.
The SOHO/MDI magnetograms \citep{Scherrer_etal95} were individual or 5-minutes averages obtained every 96 minutes except for a data gap (the SOHO summer vacation) from June to October 1998.
The SDO/HMI magnetograms \citep{Scherrer_etal12} were averaged over 12-minutes and obtained every 60 minutes.

\begin{figure}[ht]
\centerline{\includegraphics[width=\columnwidth]{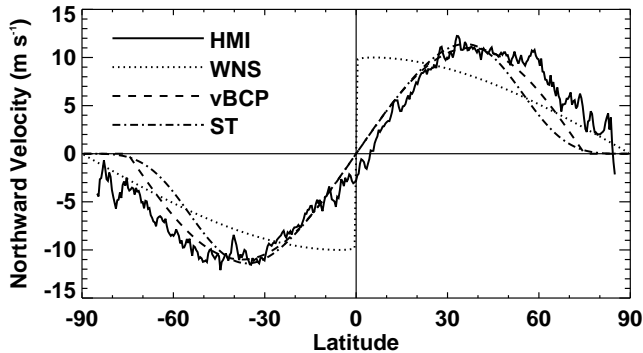}}
\caption{The average meridional flow profile obtained with data from the SDO/HMI instrument acquired between April 2010 and July 2013.
Meridional flow profiles used in previous surface flux transport models are shown with the dotted \citep{Wang_etal89}, dashed \citep{vanBallegooijen_etal98} and dashed-dotted \citep{SchrijverTitle01} lines.}
\label{fig:MFprofile}
\end{figure}

We projected each full-disk magnetogram onto a Mercator projection grid in heliographic longitude and latitude.
We extracted long, thin strips of data with longitudinal widths of $105^\circ$ and latitudinal heights of $2^\circ$ centered on the central meridian at a series of latitudes.
These strips were cross-correlated with similar strips from magnetograms obtained 8 hours later and 8 hours earlier.
The offset in longitude and latitude giving the highest correlation gives the differential rotation and meridional flow.
The average meridional flow profile obtained with the SDO/HMI data is shown in Fig.~\ref{fig:MFprofile}.

This meridional flow profile is problematic for most surface flux transport models -- it indicates that the poleward motions of the magnetic elements peak in mid-latitudes and extend right to the poles.
The early surface flux transport models \citep{Wang_etal89} employed a meridional flow that had maximum poleward velocities immediately adjacent to the equator with a slow fall-off to higher latitudes in each hemisphere (dotted line in Fig.~\ref{fig:MFprofile}.).
Later models \citep{vanBallegooijen_etal98, SchrijverTitle01} used profiles that do match ours up to latitudes of $40-50^\circ$ but then have little or no poleward flow at latitudes above $75^\circ$ (dashed and dashed-dotted lines in Fig.~\ref{fig:MFprofile}).

The latitudinal structure of the meridional flow can significantly alter the strength and structure of the polar fields it produces in these models.
The continued poleward flow at high latitudes tends to make the fields more tightly concentrated at the poles.
A peak flow velocity at very low latitudes keeps opposite polarity elements from canceling across the equator.

\section{Meridional Flow Speed Variations}

We calculated the average profile for each 27-day rotation of the Sun and fit each of them to orthogonal polynomials in $\sin B$ where $B$ is the heliographic latitude.
The fits to the meridional flow are dominated by the terms of order 1 and 3.

\begin{figure}[ht]
\centerline{\includegraphics[width=\columnwidth]{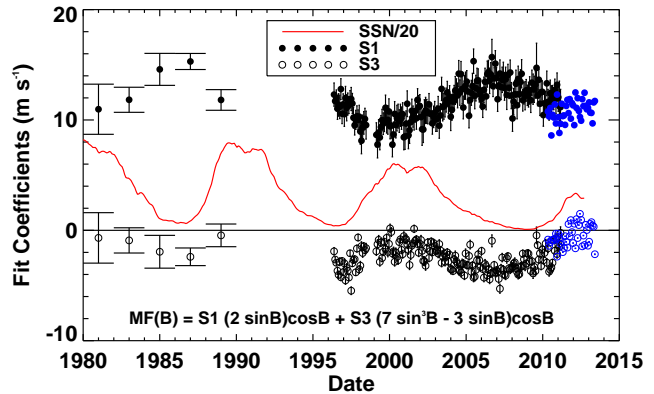}}
\caption{The meridional flow speed history from 1980 through mid-2013 is represented in terms of the coefficients (S1 and S3) of the polynomials fit to the profiles.
Results for individual Carrington rotations are shown in black for data from SOHO/MDI and in blue for the new results from SDO/HMI.
Both sets of data are shown with $2\sigma$ error bars.
The data points (with $1\sigma$ error bars) for the 2-year intervals from 1980 to 1990 from \citet{Komm_etal93B} are also shown in black.
The smoothed sunspot number is shown in red as a reference for comparing the meridional flow speed variations with the phases of the sunspot cycles.
}
\label{fig:MFspeedHistory}
\end{figure}

Fig.~\ref{fig:MFspeedHistory} shows the history of the fit coefficients for the meridional flow profiles.
Here the coefficients found with SOHO/MDI for cycle 23 \citep{HathawayRightmire10} are now augmented by those we find with SDO/HMI data for cycle 24 and by those found with NSO/Kitt Peak data for cycles 21 and 22 by \citet{Komm_etal93B} using a similar method.
(\citet{Komm_etal93B} correlated square areas at $\sim 24$ hour time lags.
We both excluded areas of strong field associated with sunspots.)
The variations in the meridional flow speed over the course of each cycle are substantial.
The meridional flow is fast at cycle minima and slow at cycle maxima as was previously noted by \citet{Komm_etal93B}.

Fig.~\ref{fig:MFspeedHistory} also shows variations in the meridional flow speed from cycle-to-cycle.
The meridional flow was slower at both the preceding minimum (1996) and the maximum (2000) of cycle 23 when compared to the minima and maxima of both the earlier cycles (21 and 22) and the later cycle (24).

\section{Meridional Flow Structure Variations}

\begin{figure}[ht]
\centerline{\includegraphics[width=\columnwidth]{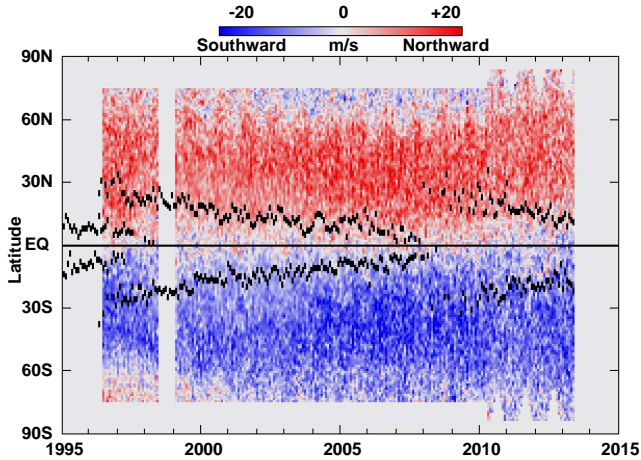}}
\caption{The meridional flow profile history from May 1996 to July 2013 is represented by showing the flow profiles from each 27-day solar rotation as color-coded vertical strips.
Shades of blue represent southward flow while shades of red represent northward flow.
Black dots indicate the location of the active latitudes as given by the latitudinal centroid of the sunspot group area in each hemisphere.}
\label{fig:MFprofileHistory}
\end{figure}

The fit coefficients shown in Fig.~\ref{fig:MFspeedHistory} help to characterize the amplitude of the variability but do not fully describe the meridional flow variations.
\citet{CameronSchussler10} noted that the variations in the S1 term shown in Fig.~\ref{fig:MFspeedHistory} could be explained in terms of the previously reported
changes to the meridional flow profiles derived from local helioseismology \citep{ZhaoKosovichev04, Gizon04, GonzalezHernandez_etal08, GonzalezHernandez_etal10} and interpreted as inflows toward the active latitudes.
This behavior is seen in Fig.~\ref{fig:MFprofileHistory} by showing each individual profile in a color-coded image.

The nature of the slow-down of the meridional flow at cycle maxima is clearly evident in Fig~\ref{fig:MFprofileHistory}.
The slow-down is seen as a weakening of the poleward flow on the poleward sides of the active latitudes (lighter shades of red in the north and blue in the south.
This can indeed be interpreted as an inflow toward the active latitudes that is superimposed on the average poleward flow profile as suggested by \citet{CameronSchussler10}, noted earlier from helioseismology \citep{ZhaoKosovichev04, Gizon04, GonzalezHernandez_etal08, GonzalezHernandez_etal10}, and from one of our earlier studies of magnetic element motion \citep{HathawayRightmire11}.

Fig.~\ref{fig:MFprofileHistory} also shows substantial differences between cycle 23 and our new results for cycle 24.
The weakening of the meridional flow on the poleward sides of the active latitudes is hardly discernible in cycle 24.
We also find polar counter-cells (equatorward flow at high latitudes) with the MDI data during cycle 23 that are not yet seen in HMI data during cycle 24.
Cycle 23 appeared to have a counter-cell in the south that extended down to $60^\circ$ in mid-1996.
By 2001 the counter-cell boundary moved poleward of our $75^\circ$ observation limit.
At about this time a counter-cell is seen to dip below this limit in the north and then proceed to grow to its maximum extent in 2006-2007 after which it shrinks but is still apparent in MDI data in 2010.
It is worth noting that during the 1-year overlap between MDI and HMI from April of 2010 to March of 2011 this northern counter-cell was visible in the MDI data but not seen in the HMI data even at $85^\circ$ north \citep{RightmireUpton_etal12}.

\section{Possible Effects on the Polar Fields}

The effects of these meridional flow variations on the polar fields and the axial dipole need to be determined.
Several previous studies have characterized the effects of simply changing the amplitude of the meridional flow.
In addition, \citet{Jiang_etal10} and \citet{CameronSchussler12} have recently examined the effects associated with variations in the shape of the meridional flow profile that are very similar to those found here (inflows toward the active latitudes).

The parametric study of \citet{Baumann_etal04} used the meridional flow profile of \citet{vanBallegooijen_etal98} (dashed line in Fig.~\ref{fig:MFprofile}) and varied the peak velocity from 0 to 30 m s$^{-1}$ with a fixed diffusivity of 600 km$^2$ s$^{-1}$.
They found that the polar fields increased in strength as the meridional flow amplitude increased from 0 to 8 m s$^{-1}$ --- more high latitude, following polarity flux is carried to the poles and the meridional flow itself counters the diffusion away from the poles.
As the meridional flow amplitude increases above 8 m s$^{-1}$ the polar fields become weaker  --- the higher velocities prevent opposite polarities from canceling across the equator so that the net flux carried by the meridional flow decreases.
(This result is similar to what was found earlier by \citet{Wang_etal89}).

The switchover in the sensitivity of polar field strength to meridional flow speed variations at 8 m s$^{-1}$ makes any conclusions about the effects of the observed variations difficult based on this study.
The average flow speed of 11 m s$^{-1}$ is very close to this switchover point and the switchover point itself must depend on the diffusivity and meridional flow profile used in the calculations.

The structural changes to the meridional flow profile (the slowdown above the active latitudes) should have a direct and negative effect on the poleward transport of the high latitude following polarity but have little effect on the flux cancellation across the equator.
In general, this should modulate the sunspot cycle amplitudes -- big cycles would have stronger inflows that would produce weaker polar fields than expected from the increase of active region sources.
In fact, we find (Fig.~\ref{fig:MFprofileHistory}) that this inflow was stronger in cycle 23 than it is in cycle 24.
\citet{Jiang_etal10} and \citet{CameronSchussler12} show that variations like this may indeed provide a nonlinear feedback on the amplitudes of the solar cycles and may even be a key ingredient in determining the amplitudes of solar cycles.
Although the variations they employ are significantly stronger than those we find in cycles 23 and 24 (neither cycle shows any actual equatorward flow within the active latitudes) their results do suggest that the meridional flow slowdown poleward of the active latitudes in cycle 23 contributed to producing the weak polar fields.

\section{Conclusions}

The exceptional depth of cycle 23/24 minimum and the exceptional length of cycle 23 can both be attributed to the small amplitude of cycle 24 -- small cycles typically start late and leave behind low minima \citep{Hathaway_etal99}.
The small amplitude of cycle 24 can be attributed to the weak polar fields produced prior to cycle 23/24 minimum by the activity and flux transport of cycle 23 -- weak polar fields produce weak cycles \citep{MunozJaramillo_etal13A}.
The strength of the polar fields (or the axial dipole moment) is largely determined by a combination of three processes: the total magnetic flux emerging in active regions, the characteristic tilt of those active regions, and the transport (meridional flow and diffusion) of that emerging flux.

Cycle 23 was significantly smaller than cycles 21 and 22 that preceded it.
This change in active region sources alone can be a significant source of the change in polar fields.
\citet{DasiEspuig_etal10} found that there are changes to the characteristic tilt of active regions as a function of cycle amplitude.
However, they did not examine cycle 23 for comparison with other cycles.
Here we examined changes in the meridional flow measured from the motions of the small magnetic elements that populate the Sun's surface.
We found variations in the meridional flow that may have contributed to producing the weak polar fields in cycle 23 and may play a more general role in modulating the amplitudes of the solar cycle.

We found that cycle 23 was characterized by a slower meridional flow and that this variation in the meridional flow was primarily due to a slowdown of the meridional flow at the latitudes poleward of the sunspot zones.
This slowdown is not seen in cycle 24 -- a much weaker solar cycle.
This suggests that this is a characteristic feature of solar cycles: the weakening of the meridional flow is greater in bigger cycles.
\citet{CameronSchussler12} suggest that this may be a key nonlinear process needed to modulate the amplitudes of the solar cycles.

We suggest that this change in the meridional flow contributed to the production of the weak polar fields at cycle 23/24 minimum.
While this suggestion is supported by the modeling work of \citet{CameronSchussler12}, more detailed modeling work with the actual meridional flow profiles is needed for confirmation.

%%% End of body of article:

%%%%%%%%%%%%%%%%%%%%%%%%%%%%%%%
%% Appendix
%
\appendix
\section{Flow Measurement Methods}

Measurements of the meridional flow can be made by a variety of techniques including direct Doppler, time-distance helioseismology, and magnetic feature tracking.
The different techniques give information about flows at different depths and about motions of different features.
Both the direct Doppler and time-distance helioseismology methods must first characterize and remove systematic signals larger than the meridional flow signal itself.
Magnetic feature tracking is not subject to such large systematic errors and has the distinct advantage of measuring the motions of the features of interest for magnetic flux transport -- magnetic field elements.

\begin{figure}[ht!]
\centerline{\includegraphics[width=\columnwidth]{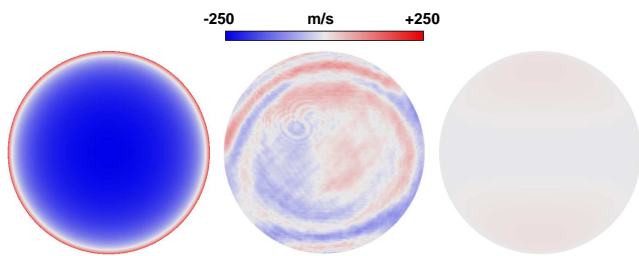}}
\caption{The direct Doppler signals associated with measuring the meridional flow are shown here with the same scaling.
The convective Blue Shift (left) has a dynamic range of 500 m s$^{-1}$ and is show here without any local variations due to the presence of magnetic field.
Imaging artifacts (center, from HMI) can have a dynamic range of 200 m s$^{-1}$.
The meridional flow (right) has a dynamic range of only 10-15 m s$^{-1}$.}
\label{fig:MFdopplerSignal}
\end{figure}

The Doppler signal due to the axisymmetric meridional flow is masked by the much larger signal due to the convective blue shift and by instrumental/scanning artifacts.
The convective blue shift is produced by the correlation between brightness and radial flow velocities in granules.
It gives a blue shift at disk center that falls off toward the limb and, depending on spectral line, becomes a red shift near the limb itself.
The convective blue shift signal typically varies by $\sim500$ m s$^{-1}$ from disk center to limb and the signal is known to vary locally in the presence of magnetic field elements.
Various methods have been devised to separated the convective blue shift signal from the meridional flow signal \citep{Snodgrass84, Ulrich_etal88, Hathaway96} but all require an accurate measurement of the convective blue shift and all assume that this signal does not vary with latitude.
Imaging artifacts, which (like the convective blue shift and meridional flow signals) remain relatively fixed on the image can also introduce systematic errors to the measurement of the meridional flow.
Fig.~\ref{fig:MFdopplerSignal} shows examples of these Doppler signals to illustrate the difficulties entailed in making direct Doppler measurements of the meridional flow.
While this method is sensitive to the meridional flow at high latitudes, it is insensitive to the meridional flow near the equator.

Local helioseismology measures the meridional flow by determining small differences in the sound travel time between north-to-south and south-to-north moving acoustic waves.
Time-Distance helioseismology does this by correlating signals observed at points separated in latitude.
\citet{DuvallHanasoge09} recently reported on a previously unnoticed systematic center-to-limb variation in acoustic wave travel times.
The source of this signal is still uncertain but it is clearly seen as an apparent flow away from disk center that has affected all previous measurements of the meridional flow using this method.
The signal itself can be characterized by measuring the signal away from disk center along the equator as was done by \citet{Zhao_etal12}.
Their characterization of this signal indicates that it can be several times larger than the meridional flow signal itself (see their Fig. 2.).
As with the direct Doppler method, this method requires an accurate measurement of a systematic signal and assumes that this signal does not vary with latitude.
This method \textit{is} sensitive to meridional flow near the equator and it provides important depth information but it becomes more uncertain at high latitudes.
Even with high resolution data these measurements rarely extend poleward of $60^\circ$.
Furthermore, when using meridional flow profiles from helioseismology one must chose the appropriate depth for the flux transport profile.

The feature tracking method using the small magnetic elements has significant advantages over either direct Doppler or local helioseismology.
It can be used to measure the meridional flow from the equator up to at least $85^\circ$ latitude \citep{RightmireUpton_etal12}.
It does not require characterizing and removing any substantial systematic signals and
it directly measures the motions of the features of interest for magnetic flux transport -- the magnetic elements.
Conversely, the motions of the magnetic elements are not given by the motions of the surface plasma.
The rotation rate of the small magnetic elements \citep{Snodgrass83, Komm_etal93A} is significantly faster than the surface Doppler rate -- indicating that the magnetic elements are anchored further down in the surface shear layer.

We should note that \citet{Dikpati_etal10A} have suggested that magnetic element feature tracking \textit{is} subject to a large systematic signal associated with the ``diffusion'' produced by supergranules.
They argue that the diffusive transport of magnetic elements away from the active latitudes should give a fictitious flow velocity away from the active latitudes with this method and they estimated the magnitude of the effect using a 1D (latitudinal) transport model in which there were no magnetic elements \textit{per se}.
We \citep{HathawayRightmire11} investigated this possibility by transporting magnetic elements in both latitude and longitude using \textit{only} supergranular flows and then attempted to measure this associated fictitious meridional flow.
We found that it must be less than the 1-2 m s$^{-1}$ noise level in our measurements (Fig. 10 of that paper).
Furthermore, the meridional flow variations we find using magnetic feature tracking are in the form of an \textit{inflow} toward the active latitudes - not the \textit{outflow} that would be produced according to \citet{Dikpati_etal10A}.

%%%%%%%%%%%%%%%%%%%%%%%%%%%%%%%%%%%%%%%%%%%%%%%%%%%%%%%%%%%%%%%%%
%
%  ACKNOWLEDGMENTS

\begin{acknowledgments}
The authors were supported by a grant from the NASA Living with a Star Program to Marshall Space Flight Center. 
The HMI data used are courtesy of the NASA/SDO and the HMI science team.
The SOHO/MDI project was supported by NASA grant NAG5-10483 to Stanford University.
SOHO is a project of international cooperation between ESA and NASA.
\end{acknowledgments}

%% ------------------------------------------------------------------------ %%
%%  REFERENCE LIST AND TEXT CITATIONS
%

%\bibliography{Hathaway}

\begin{thebibliography}{46}
\providecommand{\natexlab}[1]{#1}
\expandafter\ifx\csname urlstyle\endcsname\relax
  \providecommand{\doi}[1]{doi:\discretionary{}{}{}#1}\else
  \providecommand{\doi}{doi:\discretionary{}{}{}\begingroup
  \urlstyle{rm}\Url}\fi

\bibitem[{\textit{{Babcock}}(1961)}]{Babcock61}
{Babcock}, H.~W. (1961), {The Topology of the Sun's Magnetic Field and the
  22-YEAR Cycle.}, \textit{\apj}, \textit{133}, 572, \doi{10.1086/147060}.

\bibitem[{\textit{{Baumann} et~al.}(2004)\textit{{Baumann}, {Schmitt},
  {Sch{\"u}ssler}, and {Solanki}}}]{Baumann_etal04}
{Baumann}, I., D.~{Schmitt}, M.~{Sch{\"u}ssler}, and S.~K. {Solanki} (2004),
  {Evolution of the large-scale magnetic field on the solar surface: A
  parameter study}, \textit{\aap}, \textit{426}, 1075--1091,
  \doi{10.1051/0004-6361:20048024}.

\bibitem[{\textit{{Cameron} and {Sch{\"u}ssler}}(2010)}]{CameronSchussler10}
{Cameron}, R.~H., and M.~{Sch{\"u}ssler} (2010), {Changes of the Solar
  Meridional Velocity Profile During Cycle 23 Explained by Flows Toward the
  Activity Belts}, \textit{\apj}, \textit{720}, 1030--1032,
  \doi{10.1088/0004-637X/720/2/1030}.

\bibitem[{\textit{{Cameron} and {Sch{\"u}ssler}}(2012)}]{CameronSchussler12}
{Cameron}, R.~H., and M.~{Sch{\"u}ssler} (2012), {Are the strengths of solar
  cycles determined by converging flows towards the activity belts?},
  \textit{\aap}, \textit{548}, A57, \doi{10.1051/0004-6361/201219914}.

\bibitem[{\textit{{Cameron} et~al.}(2010)\textit{{Cameron}, {Jiang}, {Schmitt},
  and {Sch{\"u}ssler}}}]{Cameron_etal10}
{Cameron}, R.~H., J.~{Jiang}, D.~{Schmitt}, and M.~{Sch{\"u}ssler} (2010),
  {Surface Flux Transport Modeling for Solar Cycles 15-21: Effects of
  Cycle-Dependent Tilt Angles of Sunspot Groups}, \textit{\apj}, \textit{719},
  264--270, \doi{10.1088/0004-637X/719/1/264}.

\bibitem[{\textit{{Dasi-Espuig} et~al.}(2010)\textit{{Dasi-Espuig}, {Solanki},
  {Krivova}, {Cameron}, and {Pe{\~n}uela}}}]{DasiEspuig_etal10}
{Dasi-Espuig}, M., S.~K. {Solanki}, N.~A. {Krivova}, R.~{Cameron}, and
  T.~{Pe{\~n}uela} (2010), {Sunspot group tilt angles and the strength of the
  solar cycle}, \textit{\aap}, \textit{518}, A7,
  \doi{10.1051/0004-6361/201014301}.

\bibitem[{\textit{{DeVore} et~al.}(1984)\textit{{DeVore}, {Boris}, and
  {Sheeley}}}]{DeVore_etal84}
{DeVore}, C.~R., J.~P. {Boris}, and N.~R. {Sheeley}, Jr. (1984), {The
  concentration of the large-scale solar magnetic field by a meridional surface
  flow}, \textit{Solar Phys.}, \textit{92}, 1--14, \doi{10.1007/BF00157230}.

\bibitem[{\textit{{Dikpati} et~al.}(2010)\textit{{Dikpati}, {Gilman}, and
  {Ulrich}}}]{Dikpati_etal10A}
{Dikpati}, M., P.~A. {Gilman}, and R.~K. {Ulrich} (2010), {Physical Origin of
  Differences Among Various Measures of Solar Meridional Circulation},
  \textit{\apj}, \textit{722}, 774--778, \doi{10.1088/0004-637X/722/1/774}.

\bibitem[{\textit{{Duvall} and {Hanasoge}}(2009)}]{DuvallHanasoge09}
{Duvall}, T.~L., Jr., and S.~M. {Hanasoge} (2009), {Measuring Meridional
  Circulation in the Sun}, in \textit{Solar-Stellar Dynamos as Revealed by
  Helio- and Asteroseismology: GONG 2008/SOHO 21}, \textit{Astronomical Society
  of the Pacific Conference Series}, vol. 416, edited by M.~{Dikpati},
  T.~{Arentoft}, I.~{Gonz{\'a}lez Hern{\'a}ndez}, C.~{Lindsey}, and F.~{Hill},
  p. 103.

\bibitem[{\textit{{Gizon}}(2004)}]{Gizon04}
{Gizon}, L. (2004), {Helioseismology of Time-Varying Flows Through The Solar
  Cycle}, \textit{Solar Phys.}, \textit{224}, 217--228,
  \doi{10.1007/s11207-005-4983-9}.

\bibitem[{\textit{{Gleissberg}}(1939)}]{Gleissberg39}
{Gleissberg}, W. (1939), {A long-periodic fluctuation of the sun-spot numbers},
  \textit{The Observatory}, \textit{62}, 158--159.

\bibitem[{\textit{{Gonz{\'a}lez Hern{\'a}ndez}
  et~al.}(2008)\textit{{Gonz{\'a}lez Hern{\'a}ndez}, {Kholikov}, {Hill},
  {Howe}, and {Komm}}}]{GonzalezHernandez_etal08}
{Gonz{\'a}lez Hern{\'a}ndez}, I., S.~{Kholikov}, F.~{Hill}, R.~{Howe}, and
  R.~{Komm} (2008), {Subsurface Meridional Circulation in the Active Belts},
  \textit{Solar Phys.}, \textit{252}, 235--245,
  \doi{10.1007/s11207-008-9264-y}.

\bibitem[{\textit{{Gonz{\'a}lez Hern{\'a}ndez}
  et~al.}(2010)\textit{{Gonz{\'a}lez Hern{\'a}ndez}, {Howe}, {Komm}, and
  {Hill}}}]{GonzalezHernandez_etal10}
{Gonz{\'a}lez Hern{\'a}ndez}, I., R.~{Howe}, R.~{Komm}, and F.~{Hill} (2010),
  {Meridional Circulation During the Extended Solar Minimum: Another Component
  of the Torsional Oscillation?}, \textit{\apj}, \textit{713}, L16--L20,
  \doi{10.1088/2041-8205/713/1/L16}.

\bibitem[{\textit{{Hagenaar} et~al.}(1999)\textit{{Hagenaar}, {Schrijver},
  {Title}, and {Shine}}}]{Hagenaar_etal99}
{Hagenaar}, H.~J., C.~J. {Schrijver}, A.~M. {Title}, and R.~A. {Shine} (1999),
  {Dispersal of Magnetic Flux in the Quiet Solar Photosphere}, \textit{\apj},
  \textit{511}, 932--944, \doi{10.1086/306691}.

\bibitem[{\textit{{Hale} et~al.}(1919)\textit{{Hale}, {Ellerman}, {Nicholson},
  and {Joy}}}]{Hale_etal19}
{Hale}, G.~E., F.~{Ellerman}, S.~B. {Nicholson}, and A.~H. {Joy} (1919), {The
  Magnetic Polarity of Sun-Spots}, \textit{\apj}, \textit{49}, 153,
  \doi{10.1086/142452}.

\bibitem[{\textit{{Hathaway}}(1996)}]{Hathaway96}
{Hathaway}, D.~H. (1996), {Doppler Measurements of the Sun's Meridional Flow},
  \textit{\apj}, \textit{460}, 1027, \doi{10.1086/177029}.

\bibitem[{\textit{{Hathaway} and {Rightmire}}(2010)}]{HathawayRightmire10}
{Hathaway}, D.~H., and L.~{Rightmire} (2010), {Variations in the Sun's
  Meridional Flow over a Solar Cycle}, \textit{Science}, \textit{327}, 1350--,
  \doi{10.1126/science.1181990}.

\bibitem[{\textit{{Hathaway} and {Rightmire}}(2011)}]{HathawayRightmire11}
{Hathaway}, D.~H., and L.~{Rightmire} (2011), {Variations in the Axisymmetric
  Transport of Magnetic Elements on the Sun: 1996-2010}, \textit{\apj},
  \textit{729}, 80, \doi{10.1088/0004-637X/729/2/80}.

\bibitem[{\textit{{Hathaway} et~al.}(1994)\textit{{Hathaway}, {Wilson}, and
  {Reichmann}}}]{Hathaway_etal94}
{Hathaway}, D.~H., R.~M. {Wilson}, and E.~J. {Reichmann} (1994), {The shape of
  the sunspot cycle}, \textit{Solar Phys.}, \textit{151}, 177--190,
  \doi{10.1007/BF00654090}.

\bibitem[{\textit{{Hathaway} et~al.}(1999)\textit{{Hathaway}, {Wilson}, and
  {Reichmann}}}]{Hathaway_etal99}
{Hathaway}, D.~H., R.~M. {Wilson}, and E.~J. {Reichmann} (1999), {A Synthesis
  of Solar Cycle Prediction Techniques}, \textit{\jgr}, \textit{104}, 22,375,
  \doi{10.1029/1999JA900313}.

\bibitem[{\textit{{Jiang} et~al.}(2010)\textit{{Jiang}, {I{\c s}ik}, {Cameron},
  {Schmitt}, and {Sch{\"u}ssler}}}]{Jiang_etal10}
{Jiang}, J., E.~{I{\c s}ik}, R.~H. {Cameron}, D.~{Schmitt}, and
  M.~{Sch{\"u}ssler} (2010), {The Effect of Activity-related Meridional Flow
  Modulation on the Strength of the Solar Polar Magnetic Field}, \textit{\apj},
  \textit{717}, 597--602, \doi{10.1088/0004-637X/717/1/597}.

\bibitem[{\textit{{Jiang} et~al.}(2011)\textit{{Jiang}, {Cameron}, {Schmitt},
  and {Sch{\"u}ssler}}}]{Jiang_etal11A}
{Jiang}, J., R.~H. {Cameron}, D.~{Schmitt}, and M.~{Sch{\"u}ssler} (2011), {The
  solar magnetic field since 1700. I. Characteristics of sunspot group
  emergence and reconstruction of the butterfly diagram}, \textit{\aap},
  \textit{528}, A82, \doi{10.1051/0004-6361/201016167}.

\bibitem[{\textit{{Komm} et~al.}(1993{\natexlab{a}})\textit{{Komm}, {Howard},
  and {Harvey}}}]{Komm_etal93B}
{Komm}, R.~W., R.~F. {Howard}, and J.~W. {Harvey} (1993{\natexlab{a}}),
  {Meridional Flow of Small Photospheric Magnetic Features}, \textit{Solar
  Phys.}, \textit{147}, 207--223, \doi{10.1007/BF00690713}.

\bibitem[{\textit{{Komm} et~al.}(1993{\natexlab{b}})\textit{{Komm}, {Howard},
  and {Harvey}}}]{Komm_etal93A}
{Komm}, R.~W., R.~F. {Howard}, and J.~W. {Harvey} (1993{\natexlab{b}}),
  {Rotation rates of small magnetic features from two- and one-dimensional
  cross-correlation analyses}, \textit{Solar Phys.}, \textit{145}, 1--10,
  \doi{10.1007/BF00627979}.

\bibitem[{\textit{{Leighton}}(1969)}]{Leighton69}
{Leighton}, R.~B. (1969), {A Magneto-Kinematic Model of the Solar Cycle},
  \textit{\apj}, \textit{156}, 1, \doi{10.1086/149943}.

\bibitem[{\textit{{Mu{\~n}oz-Jaramillo}
  et~al.}(2013)\textit{{Mu{\~n}oz-Jaramillo}, {Dasi-Espuig}, {Balmaceda}, and
  {DeLuca}}}]{MunozJaramillo_etal13A}
{Mu{\~n}oz-Jaramillo}, A., M.~{Dasi-Espuig}, L.~A. {Balmaceda}, and E.~E.
  {DeLuca} (2013), {Solar Cycle Propagation, Memory, and Prediction: Insights
  from a Century of Magnetic Proxies}, \textit{\apjl}, \textit{767}, L25,
  \doi{10.1088/2041-8205/767/2/L25}.

\bibitem[{\textit{{Ohl}}(1966)}]{Ohl66}
{Ohl}, A.~I. (1966), {Forecast of sunspot maximum number of cycle 20},
  \textit{Byulletin Solnechnye Dannye Akademie Nauk SSSR}, \textit{9}, 84.

\bibitem[{\textit{{Rightmire-Upton} et~al.}(2012)\textit{{Rightmire-Upton},
  {Hathaway}, and {Kosak}}}]{RightmireUpton_etal12}
{Rightmire-Upton}, L., D.~H. {Hathaway}, and K.~{Kosak} (2012), {Measurements
  of the Sun's High-latitude Meridional Circulation}, \textit{\apj},
  \textit{761}, L14, \doi{10.1088/2041-8205/761/1/L14}.

\bibitem[{\textit{{Rouillard} et~al.}(2007)\textit{{Rouillard}, {Lockwood}, and
  {Finch}}}]{Rouillard_etal07}
{Rouillard}, A.~P., M.~{Lockwood}, and I.~{Finch} (2007), {Centennial changes
  in the solar wind speed and in the open solar flux}, \textit{Journal of
  Geophysical Research (Space Physics)}, \textit{112}, A05103,
  \doi{10.1029/2006JA012130}.

\bibitem[{\textit{{Schatten} et~al.}(1996)\textit{{Schatten}, {Myers}, and
  {Sofia}}}]{Schatten_etal96}
{Schatten}, K., D.~J. {Myers}, and S.~{Sofia} (1996), {Solar activity forecast
  for solar cycle 23}, \textit{\grl}, \textit{23}, 605--608,
  \doi{10.1029/96GL00451}.

\bibitem[{\textit{{Schatten} and {Sofia}}(1987)}]{SchattenSofia87}
{Schatten}, K.~H., and S.~{Sofia} (1987), {Forecast of an exceptionally large
  even-numbered solar cycle}, \textit{\grl}, \textit{14}, 632--635,
  \doi{10.1029/GL014i006p00632}.

\bibitem[{\textit{{Schatten} et~al.}(1978)\textit{{Schatten}, {Scherrer},
  {Svalgaard}, and {Wilcox}}}]{Schatten_etal78}
{Schatten}, K.~H., P.~H. {Scherrer}, L.~{Svalgaard}, and J.~M. {Wilcox} (1978),
  {Using dynamo theory to predict the sunspot number during solar cycle 21},
  \textit{\grl}, \textit{5}, 411--414, \doi{10.1029/GL005i005p00411}.

\bibitem[{\textit{{Scherrer} et~al.}(1995)}]{Scherrer_etal95}
{Scherrer}, P.~H., et~al. (1995), {The Solar Oscillations Investigation -
  Michelson Doppler Imager}, \textit{Solar Phys.}, \textit{162}, 129--188,
  \doi{10.1007/BF00733429}.

\bibitem[{\textit{{Scherrer} et~al.}(2012)}]{Scherrer_etal12}
{Scherrer}, P.~H., et~al. (2012), {The Helioseismic and Magnetic Imager (HMI)
  Investigation for the Solar Dynamics Observatory (SDO)}, \textit{Solar
  Phys.}, \textit{275}, 207--227, \doi{10.1007/s11207-011-9834-2}.

\bibitem[{\textit{{Schrijver} and {Title}}(2001)}]{SchrijverTitle01}
{Schrijver}, C.~J., and A.~M. {Title} (2001), {On the Formation of Polar Spots
  in Sun-like Stars}, \textit{\apj}, \textit{551}, 1099--1106,
  \doi{10.1086/320237}.

\bibitem[{\textit{{Snodgrass}}(1983)}]{Snodgrass83}
{Snodgrass}, H.~B. (1983), {Magnetic rotation of the solar photosphere},
  \textit{\apj}, \textit{270}, 288--299, \doi{10.1086/161121}.

\bibitem[{\textit{{Snodgrass}}(1984)}]{Snodgrass84}
{Snodgrass}, H.~B. (1984), {Separation of large-scale photospheric Doppler
  patterns}, \textit{Solar Phys.}, \textit{94}, 13--31,
  \doi{10.1007/BF00154804}.

\bibitem[{\textit{{Svalgaard} and {Cliver}}(2005)}]{SvalgaardCliver05}
{Svalgaard}, L., and E.~W. {Cliver} (2005), {The IDV index: Its derivation and
  use in inferring long-term variations of the interplanetary magnetic field
  strength}, \textit{Journal of Geophysical Research (Space Physics)},
  \textit{110}.

\bibitem[{\textit{{Svalgaard} et~al.}(2005)\textit{{Svalgaard}, {Cliver}, and
  {Kamide}}}]{Svalgaard_etal05}
{Svalgaard}, L., E.~W. {Cliver}, and Y.~{Kamide} (2005), {Sunspot cycle 24:
  Smallest cycle in 100 years?}, \textit{\grl}, \textit{32}, L01104,
  \doi{10.1029/2004GL021664}.

\bibitem[{\textit{{Ulrich} et~al.}(1988)\textit{{Ulrich}, {Boyden}, {Webster},
  {Padilla}, and {Snodgrass}}}]{Ulrich_etal88}
{Ulrich}, R.~K., J.~E. {Boyden}, L.~{Webster}, S.~P. {Padilla}, and H.~B.
  {Snodgrass} (1988), {Solar rotation measurements at Mount Wilson. V -
  Reanalysis of 21 years of data}, \textit{Solar Phys.}, \textit{117},
  291--328, \doi{10.1007/BF00147250}.

\bibitem[{\textit{{van Ballegooijen} et~al.}(1998)\textit{{van Ballegooijen},
  {Cartledge}, and {Priest}}}]{vanBallegooijen_etal98}
{van Ballegooijen}, A.~A., N.~P. {Cartledge}, and E.~R. {Priest} (1998),
  {Magnetic Flux Transport and the Formation of Filament Channels on the Sun},
  \textit{\apj}, \textit{501}, 866, \doi{10.1086/305823}.

\bibitem[{\textit{{Waldmeier}}(1935)}]{Waldmeier35}
{Waldmeier}, M. (1935), {Neue Eigenschaften der Sonnenfleckenkurve},
  \textit{Astronomische Mitteilungen der Eidgen{\"o}ssischen Sternwarte
  Zurich}, \textit{14}, 105--136.

\bibitem[{\textit{{Wang} and {Sheeley}}(2009)}]{WangSheeley09}
{Wang}, Y.-M., and N.~R. {Sheeley} (2009), {Understanding the Geomagnetic
  Precursor of the Solar Cycle}, \textit{\apjl}, \textit{694}, L11--L15,
  \doi{10.1088/0004-637X/694/1/L11}.

\bibitem[{\textit{{Wang} et~al.}(1989)\textit{{Wang}, {Nash}, and
  {Sheeley}}}]{Wang_etal89}
{Wang}, Y.-M., A.~G. {Nash}, and N.~R. {Sheeley}, Jr. (1989), {Evolution of the
  sun's polar fields during sunspot cycle 21 - Poleward surges and long-term
  behavior}, \textit{\apj}, \textit{347}, 529--539, \doi{10.1086/168143}.

\bibitem[{\textit{{Zhao} and {Kosovichev}}(2004)}]{ZhaoKosovichev04}
{Zhao}, J., and A.~G. {Kosovichev} (2004), {Torsional Oscillation, Meridional
  Flows, and Vorticity Inferred in the Upper Convection Zone of the Sun by
  Time-Distance Helioseismology}, \textit{\apj}, \textit{603}, 776--784,
  \doi{10.1086/381489}.

\bibitem[{\textit{{Zhao} et~al.}(2012)\textit{{Zhao}, {Nagashima}, {Bogart},
  {Kosovichev}, and {Duvall}}}]{Zhao_etal12}
{Zhao}, J., K.~{Nagashima}, R.~S. {Bogart}, A.~G. {Kosovichev}, and T.~L.
  {Duvall}, Jr. (2012), {Systematic Center-to-limb Variation in Measured
  Helioseismic Travel Times and its Effect on Inferences of Solar Interior
  Meridional Flows}, \textit{\apj}, \textit{749}, L5,
  \doi{10.1088/2041-8205/749/1/L5}.

\end{thebibliography}
%\bibliographystyle{agu08}

%% ------------------------------------------------------------------------ %%
%
%  END ARTICLE
%
%% ------------------------------------------------------------------------ %%
\end{article}

\end{document}